# Dynamical force measurements for contacting soft surfaces upon steady sliding: Fixed-depth tribology

Cheng-En Tsai [3,1,2,*] and Jih-Chiang(JC) Tsai [1,#]

[1)]Institute of Physics, Academia Sinica, Taipei, Taiwan
[2)]Department of Physics, National Central University, Taoyuan, Taiwan
[3)]Department of Physics, National Taiwan University, Taipei, Taiwan

*) d11222018@ntu.edu.tw
#) jctsai@phys.sinica.edu.tw

ABSTRACT

The tribology between surfaces can have a profound impact on the response of a mechanical system, such as how granular particles are driven to flow. In this work, we perform experiments that time-resolve the tangential and normal components of the force between two semi-cylindrical PDMS (polydimethylsiloxane) samples immersed in fluid, as they slide against each other in a range of controlled speeds. The time-averaged friction force shows a non-monotonic dependence on the sliding speed over four decades, which is consistent to the paradigmatic Stribeck diagram and three dynamical regimes associated with it. Our specially designed fixed-depth setup allows us to study the fluctuation of force that exhibits strong stick-slip patterns in one of the regimes. Data from repetitive experiments reveal that both the "onset speed" for the stick-slip patterns and its spatial location along the sample change gradually during the course of our experiments, indicating changes on the sample surfaces. In addition, we conduct counterpart experiments by using spherical samples rubbing against each other, to make a direct connection of the inter-particle tribology to the granular flow reported in our previous work [Phys. Rev. Lett. 126,128001 (2021)].

# I. Introduction

Friction between solid surfaces is ubiquitous in daily lives. There has been a long tradition for studies of the "frictional force" between sliding surfaces, also known as tribology [1,2]. While a common starting point for discussing dynamical friction is to model it with a speed-independent constant, known as Amonton's Law or Coulomb friction, it is also well known that friction force often varies with sliding speed [2–4]. For a wide range of materials, often with fluid between the contacting surfaces, how the tangential force (the "friction") between them depends on the sliding speed is often summarized as a Stribeck diagram [2,4–8] that exhibit three dynamical regimes: (1) At a constant normal load and at "the slow limit", the regime of *boundary lubrication* exhibits a plateau value for the tangential force; (2) For a range of intermediate speeds, the regime of *mixed lubrication* exhibits a decline of tangential force upon the increase of the sliding speed. (3) As the sliding speed goes further above, the tangential force rises in the *hydrodynamic regime*. The occurrence of these three regimes have been investigated in depth for various material surfaces [5–16]. For example, Stokes and colleagues measured the friction force on PDMS(polydimethylsiloxane) and PTFE(Polytetrafluoroethylene) elastomers with interstitial lubricant, revealing the non-monotonic speed dependence corresponding to the three regimes described above [5,6]; Israelachvili and coworkers time-resolves the friction force between two mica surfaces with interstitial fluid, revealing further details of dynamics in the mixed-lubrication regime through monitoring the fluctuation of force [17,18]. In addition to the driving speed, there are other factors that also influence the friction have been studied in depth, such as the roughness [19–22], effect of lubricant [6,11,23], and the hydrophilicity [24,25].

In studies of granular flows, two recent works have drawn our attention to the role of tribology that might have been underaddressed in granular systems. In Ref. [26], Workamp and Dijksman have established how the rheology of granular flows varies substantially with the dynamical friction of the constituent particles, which they characterized by direct measurements. In our recent work cited as Ref. [27], we reported an intriguing instability in a granular shear flow: The stick-slip intermittency occurs, but only in the mid-range of the shear rates explored. We speculated that the tribology in the mixed-lubrication could have played a significant role on such instability.

To characterize the essential tribology for modelling flows of tightly packed granular particles like those described in Ref. [27], we focus on the tribology of PDMS surfaces that are immersed in glycerol-water mixture. In addition, to obtain insight for the stick-slip responses in mixed-lubrication regime, we use semi-cylindrical samples with a "fixed-depth" mechanism. These aspects and the working principles are described in the next section. Section III illustrates the general behavior of the time-averaged response that is consistent to the three regimes well known in traditional fixed-load tribology. Section IV shows our time-resolved measurements of the stick-slip fluctuation in force. These fluctuations also reveal clues on the gradual change of the sample surfaces as the result of repetitive experiments. In Section V, we describe our additional experiments that utilize spherical samples, for a straightforward connection of inter-particle tribology to the granular flow.

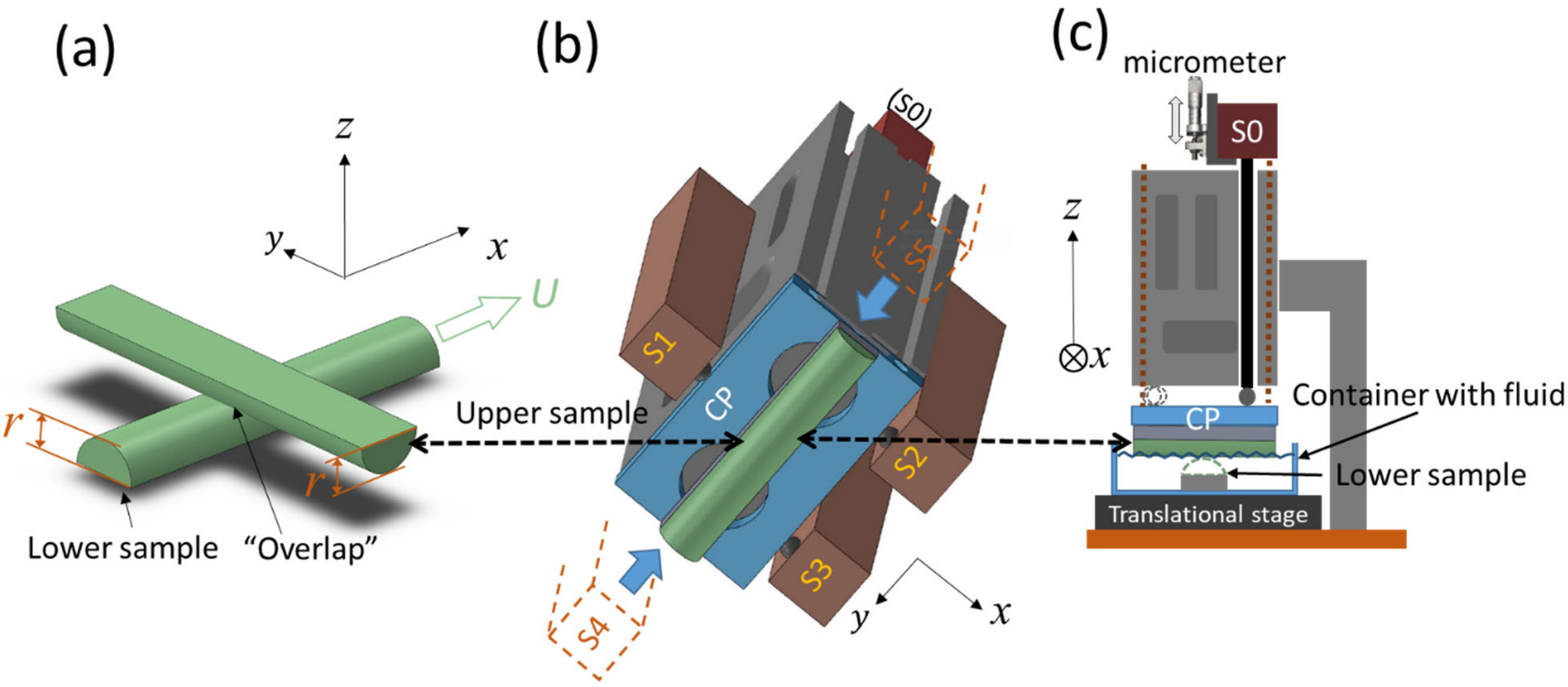

**Fig.1:** Experimental setup --- **(a)** Schematics of the two orthogonally arranged semi-cylindrical samples, with an enlarged arrow indicating the motion of the lower one, in the direction of *x*. The primamry radius of curverture of both samples is labeled as $r = 4.5$ mm. **(b)** Arrangement of the upper sample, a central piece (CP), and the the six sensors (S0~S5) for determining the force exerted from the lower sample --- see main texts for the function of individual sensors. All sensors at attached firmly to a aluminum block shown in gray. **(c)** Sideview of the entire system, but with the sensors S1-S5 omitted for clarity. The brown dashed lines represent two rubber bands that keep the CP from falling. The lower sample is fixed to a container that would be driven by a translational stage in the direction of *x*.

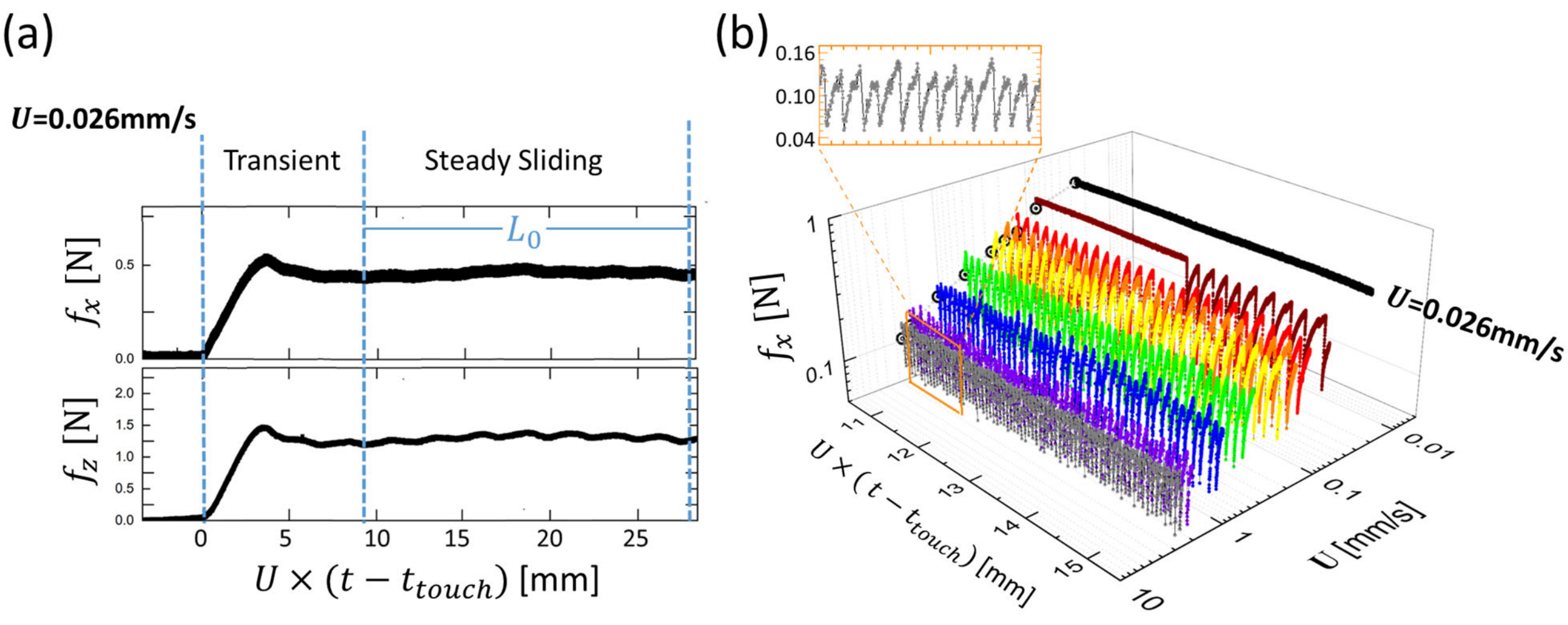

**Fig.2:** Example of raw signals. **(a)** The tangential and normal force recorded for a typical run, that produces about 20mm($L_0$) of *steady sliding* at $U = 0.026$ mm/s. The zero on the x-axis means the time that two samples touch each other. **(b)** Tangential force $f_x$ at different sliding speeds. The insect magnifies the signal at $U \sim 2.5$mm/s. For (a) and (b), $\eta = 212$ mPa-s, $D_p = 0.25mm$.

## II. Methodology

The aim of our setup is to time-resolve the force response of two semi-cylindrical samples sliding against each other. **Fig.1(a)** shows the concept: Two samples made by PDMS are arranged orthogonally and contact each other. The upper sample, along the direction of y, is 5cm-long and is attached to an assembly. The lower sample, along the direction of $x$, is 10cm-long and is attached to a container, that would be driven by a motorized translational stage at a controlled speed $U$ — to be explained further with Fig.1c. The height of the two samples are adjusted in a way that, in each run, the two semi-cylindrical surface have a constant "overlap" as shown on the graph. We refer to this overlap as the *pressing depth* $D_p$. The direction of the "pressing" defines our axis $z$, which also coincides with the direction of gravity. Extra cares are taken to make sure that the motion of the translational stage and the "ridge" of the lower sample are both aligned precisely to the $x$ axis, in order to maintain the pressing depth $D_p$ strictly constant -- see Appendix for details.

In **Fig.1(b)**, we show that the upper sample is attached to a central piece (CP), which is made of aluminum with a minimized mass to make its role on the measurement of the force fluctuations negligible. CP is held stationary by force sensors S1-S5, through steel balls to avoid undesired lateral forces. The *tangential* force, $f_x$, is determined by the signals of S1-S3. Comparing the signals of S4-S5 provides an additional check for our alignment.

**Figure 1(c)** illustrates that the lower sample situates just beneath the mid-point along the upper one. The lower sample is immersed in fluid and is driven by a motorized stage in the direction of $x$. The fluid co-moves with the lower sample during the experiments. To make its drag on the upper sample negligible, we keep the fluid level just barely above the contact point. This figure also shows that the *normal* force $f_z$ can be determined by sensor S0, through a long rod going through the aluminum block (shown in gray) that is in touch with the CP via a steel ball. Two other steel balls (shown in dotted circles) are inserted between the CP and the stationary aluminum block. The purpose of three steel balls are twofold: Firstly, they make sure that no lateral force interfere with the determination of $f_x$ (by sensors S1-S3, as described with Fig.1b). Secondly, changing the height of sensor S0 by a micrometer produces a negligible tilting of CP, which in turn fine-tunes the value of the pressing depth $D_p$. Given that the micrometer supporting S0 has a precision of 0.01mm, the value of the pressing depth $D_p$ for each experiment can be set in a three-step procedure: (1) The reading of the micrometer correponding to $D_p= 0$ is determined beforehand by lowering the upper sample incrementally until a non-zero force is detected when the two samples are in touch with each other. (2) The motorized stage drives the lower sample in the –x direction, to a stand-by position so that two samples are temporarily not in contact. (3) Adjust the reading of the micrometer, to lower the upper sample to the height corresponding to the desired value of $D_p$, before starting a standard run of our experiment.

In **Fig.2(a)**, we show the typical signals for one run of the experiment. The rise of $f_x$ and $f_z$ from zero marks the transient stage when edge of the lower sample moves gets in contact with the upper (static) one. We define the time average and study the fluctuation of force in the last 20mm (shown as $L_0$ on the graph), which we consider as the steady state. We preprogram the motorized stage so that the lower sample goes back and forth in a sequence of sliding speeds. **Fig2(b)** shows one such example, with nine different sliding speeds. In this example, significant stick-slip fluctuations emerge as the sliding speed goes above the “onset speed” at around 0.08mm/s.

The PDMS samples in our experiment are created by molding ---see Appendix for the full processes. Their Young’s modules are determined by independent experiments to be around 1.5 MPa in our previous work [27]. Through a commercial white-light interferometer (Profilm3D®), the RMS-roughness of both the mold and the sample are determined to be $(0.2 \pm 0.1)$μm.

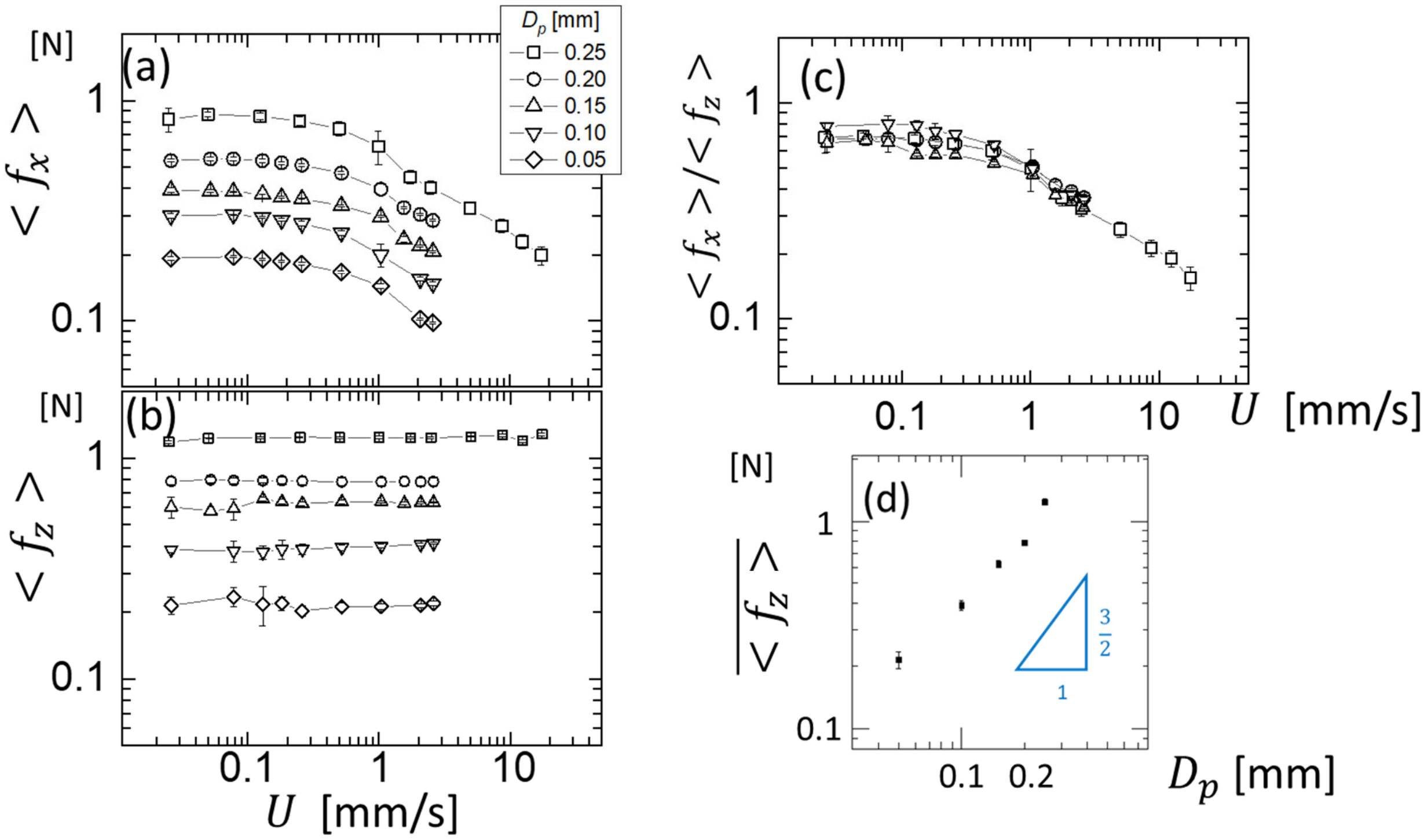

**Fig 3: (a-c)** Time-averaged force as functions of the sliding speed $U$, at different values of $D_p$: the tangential component <$f_x$>, the normal component <$f_z$>, and their ratio <$f_x$>/<$f_z$>. Each symbol represents an average over multiple runs (up to 10), with an error bar showing their standard deviations. **(d)** $\overline{< f_z >}$, the average of <$f_z$> over all sliding speeds, as a function of $D_p$. The triangle on the lower right shows the Hertzian scaling: $f_z \sim D_p^{1.5}$.

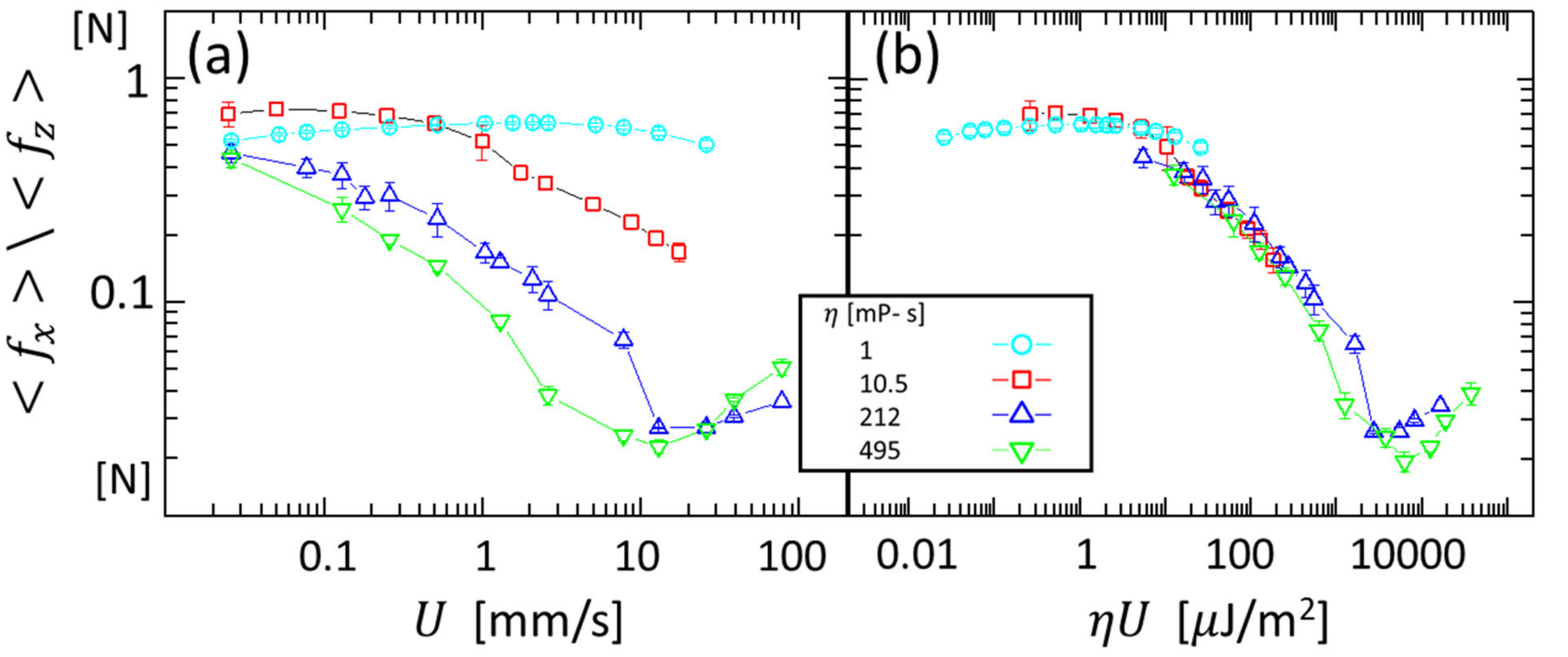

**Fig.4:** The time-averaged tangential force $< f_x >$ with 4 different interstitial fluid and the same $D_p =$ 0.25mm, normalized by the time-averaged normal force $< f_z >$. They are shown as a function of **(a)** sliding speed and **(b)** the product $\eta U$. For the two figures, the symbol represents an average over multiple runs (up to 10), with an error bar showing their standard deviations.

## III. Time-averaged Response

*A. Dependence on sliding speed and pressing depth*

The time-averaged tangential force $< f_x >$ and normal force $< f_z >$ as functions of the sliding speed $U$ are shown in **Fig.3(a-b)**, with five different values of $D_p$. Here, we utilize 60% glycerol-water mixture as the interstitial fluid, which result in 10.5 mPa-s fluid viscosity at 25°C. The sliding speed ranges from 0.026mm/s to 26mm/s for the datasets with $D_p = 0.25$ mm, and to 2.6mm/s for other values of $D_p$. For all values of $D_p$ (from 0.05 to 0.25 mm), we observe that both $< f_x >$ and $< f_z >$ exhibit a plateau at sliding speeds below 0.2 mm/s. Beyond this value, $< f_x >$ declines as $U$ increases. In contrast, the change in $< f_z >$ with the sliding speed is insignificant and appears to depend solely on the value of $D_p$. **Fig.3(c)** shows the ratio of the time-averaged tangential force and normal force $< f_x >/< f_z >$, as a function of the sliding speed**.** The data exhibit a reasonable collapse over different values of $D_p$. At low speed limit, the effective friction coefficient approaches ~0.7, but can drop to as low as ~0.35 for all datasets. In the case with $D_p = 0.25$, the friction coefficient can drop to ~0.15 at $U$ = 18 mm/s. **Fig. 3(d)** shows how the speed-independent normal force, denoted as $\overline{< f_z >}$, depends on the value of pressing depth $D_p$. The data roughly aligns with the Hertzian scaling ($f_z \sim D_p^{1.5}$) that one would anticipate from the static theories [28,29].

*B. Dependence on sliding speed and fluid viscosity*

The relationship between friction coefficient and the sliding speed of two sliding surfaces usually depends on the viscosity of interstitial fluid [5,6,8,30]. To check this effect, we adjust the concentration of our glycerol-water mixtures and obtain four different viscosities for the interstitial fluid: 1, 10.5, 212, and 486 mPa-s. **Fig. 4(a)** shows the ratio $< f_x > /< f_z >$ with the same $D_p = 0.25$mm as functions of the sliding speed. In this speed interval, the data at the lowest viscosity $\eta = 1$ mPa-s exhibits relatively small variations over the sliding speed $U$, while the data with higher viscosities $\eta = 212$ mPa-s and $495$ mPa-s show an non-monotonic relationship with the sliding speed. In **Fig. 4(b),** we replace the horizontal axis with the product $\eta U$. We find a reasonable collapse of data onto a curve resembling the paradigmatic Stribeck diagram [1–3]. The force ratio $< f_x > /< f_z >$ shows a plateau with the value around 0.6 at values of $\eta U$ smaller than 10 $\mu J/m^2$, a decline at the intermediate values, and an uprise at high values of $\eta U$. The valley of the curve reaches a value as small as 0.02. Our results are consistent with prior studies on the tribology of PDMS [8], despite the difference in sample geometries.

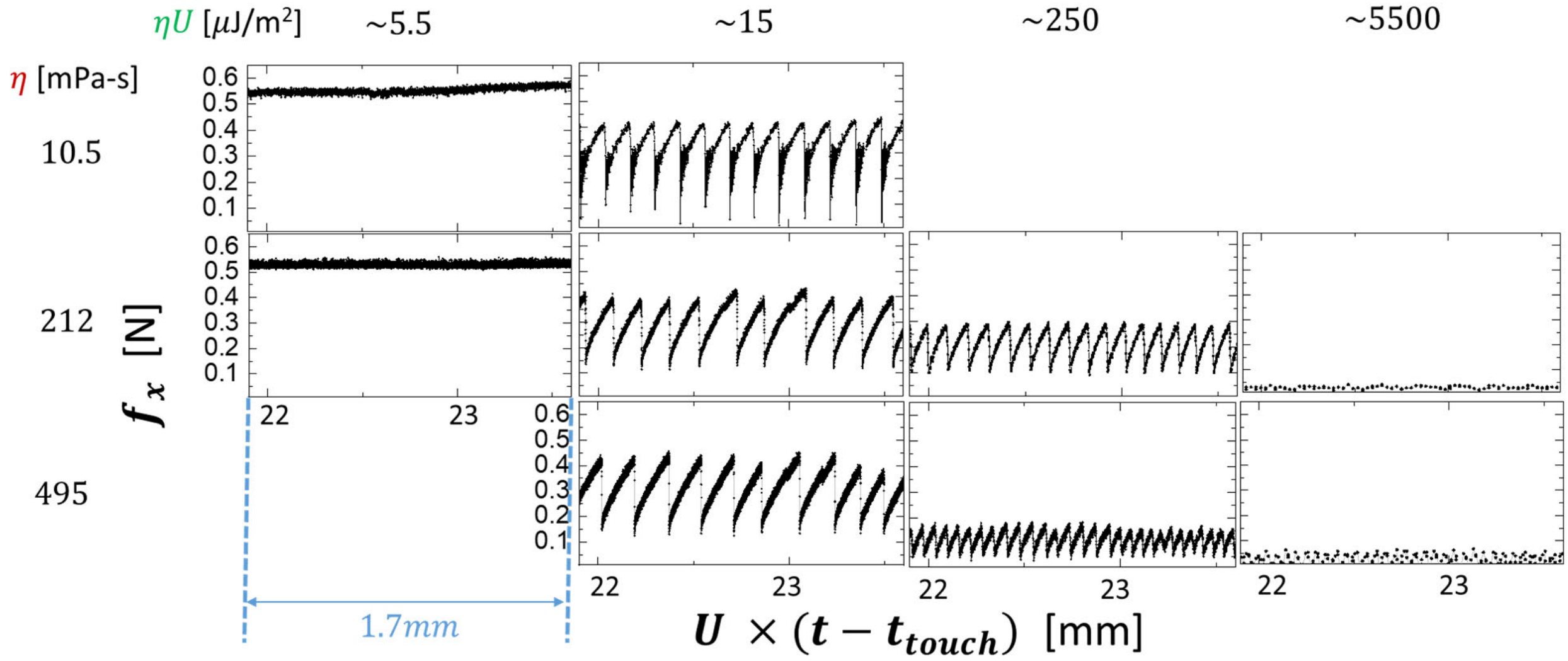

**Fig.5:** Time-series of $f_x$ plotted against the displacement $U \times t$, selected from nine different combinations of driving speed $U$ and fluid viscosity $\eta$. Panels are arranged in three columns by four rows, based on the value of U and the product $\eta U$, respectively. Time resolution of all signals are set by the same sampling rate 1500Hz.

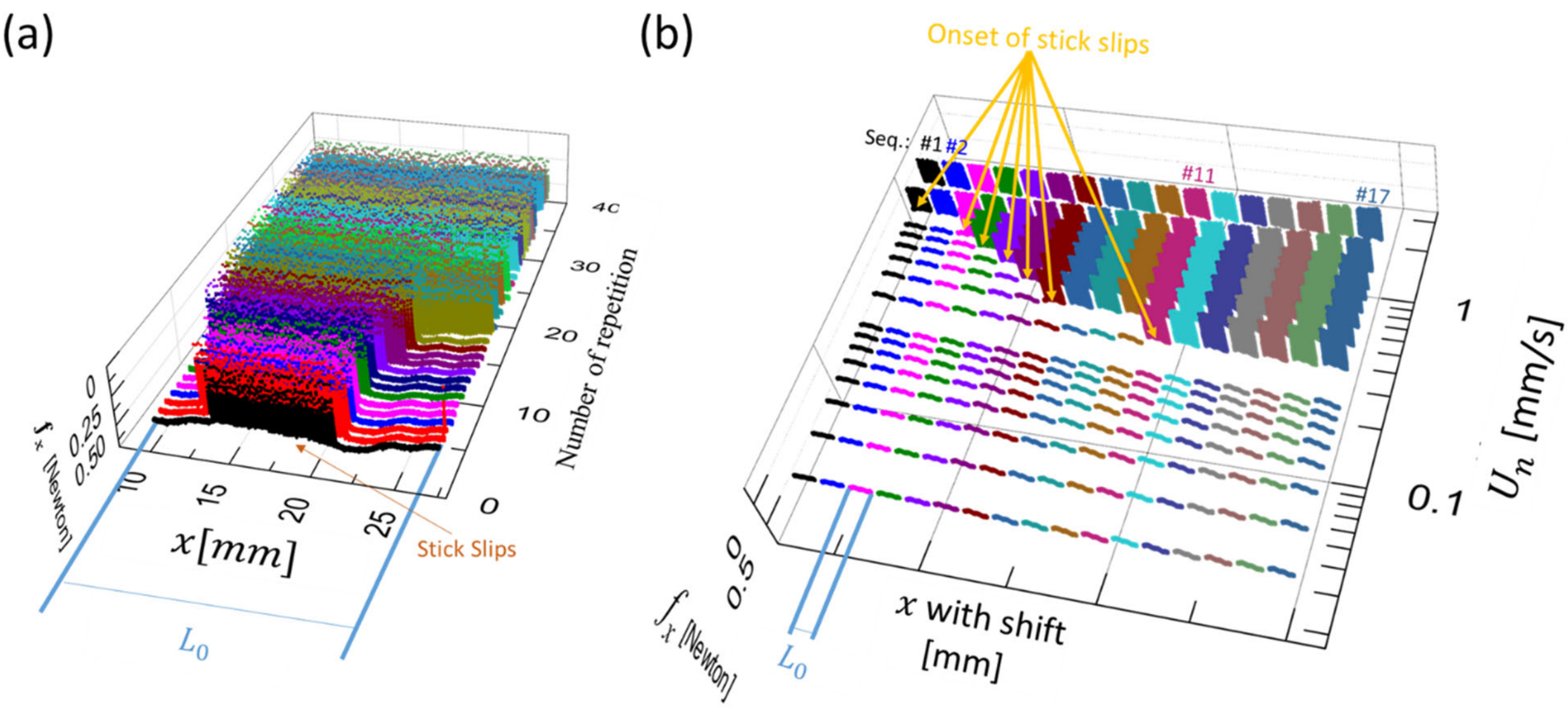

**Fig.6:** Time-series of $f_x$ plotted against the displacement $x$ for repetitive experiments. Here, unlike Fig. 2(b), value of the $f_x$ is displayed upside down for visualizing the occurrence of stick-slip pattern. $D_p = 0.25mm$ and $\eta =$ 10.5 mPa-s. $L_0$ denotes the interval of the steady state as described in Fig.2a. **(a)** Results of an experiment at a fixed sliding speed $U$=1.82mm, data for repetitive run are staked horizontally. **(b)** Results of 17 experiments, each consisting a sequence of 16 different sliding speeds $\{U_n\}$. Data of consecutive experiments are shift by a distance slightly large than $L_0$ in the direction of x, to demonstrate the "onset" of stick slips in each sequence.

# IV. Fluctuations over time and indications of surface change

*A. Typical fluctuations in tangential force and relation to prior studies*

Examples of how the tangential force $f_x$ varies over time are shown in **Fig.5.** This figure shows nine cases with different combinations of fluid viscosity $\eta$ and sliding speed $U$. For easy comparison, we place cases with similar values of $\eta U$ in one vertical column, and use the spatial displacement $U \times t$ as the horizontal axis for all subplots. Given that the data cover a wide range of $\eta$ and $U$ by a few decades, this figure shows that the occurrence of stick-slip patterns is closely related to the value of $\eta U$. Firstly, the fluctuations are insignificant for the cases with a small value of $\eta U$ ~5.5 $\mu J/m^2$. Significant stick-slip patterns are detected for cases shown by the second and the third columns. For the three data sets in the second column ($\eta U \sim 15$ $\mu J/m^2$ ), the typical interval between "spikes" differ only by ~10%, despite the difference in viscosity by a factor of ~50. The third column ($\eta U \sim 250$ $\mu J/m^2$) shows that a further increase in the value of $\eta U$ brings two outcomes: The amplitude of stick slips decreases by roughly a factor of two, and the interval between spikes decreases accordingly. Lastly, for a value of $\eta U$ as large as 5500 $\mu J/m^2$, stick slips are undetectable.

In tribology, the different patterns that we have described (in Fig.5 and Fig.2(b)) are often understood as the consequence of the system crossing the regime of *mixed lubrication* as the driving speed changes [2,4,7,23]. It is believed that the interplay of surface asperities and the interstitial fluid produces the stick-slip cycle: Some of surface asperities form direct solid-to-solid contact across interface, with the rest of the space still filled with fluid. As the two surfaces move against each other tangentially at a sufficiently high speed, the viscous stress generated by the fluid can become comparable to the elastic stress between contacting asperities. This results in substantial deformation of the contacting asperities, leading to possible "invasion" of the fluid that separates some of the contacts [6]. Once some of these deformed asperities slip, the shear stress is redistributed to the remaining contacts, making them more likely to slip as well. The movement of these slipping asperities would eventually stop, as new solid-to-solid contacts are rebuilt. This would complete a stick-slip cycle. Further increasing the driving speed can make the viscous stress much higher than the elastic force between the deformed asperities per unit area, and therefore keep the two surfaces always separated. The continuous fluid layer that keeps those asperities apart is known as the *lubrication film* and is believed to eliminate the stick-slip fluctuations, prior to the transition of *hydrodynamic lubrication*[1,2]. Such hydrodynamic transition can be seen from our data that exhibits the monotonic uprise of tangential force over the increase of $\eta U$ above the threshold value ~5500 $\mu J/m^2$ ---see Fig.4(b).

The scenario above takes into account the deformation between the contacting surfaces. On the other hand, we are aware of a recent theory by Bonn and coworkers [9] based on the lubrication dynamics between hard surfaces. For the limiting case of a sphere sliding against a plane, their theory predicts a transition from the regime of *mixed lubrication* to *hydrodynamic regime* upon $(\eta UR/F_N)^{M\to H} = 0.236 \cdot \sqrt{\sigma/R}$ , in which $R$ stands for the radius of the sphere, $\sigma$ for the surface roughness, and $F_N$ for the normal load. By substituting those variables with values in our experiment, their theory would predict a transition threshold $\eta U{\sim}4 \times 10^5$ $\mu J/m^2$ except with an order-1 factor to account for the difference in geometries. We believe that the fact that our hydrodynamic transition occurs at a much lower value of $\eta U$ than the prediction by the hard-surface theory reflects the importance of the fluid-structure interaction in the context of our experiment: The fluid drag deforms both PDMS surfaces and push them apart, thus the threshold of this hydrodynamic transition is significantly lowered than the prediction by theories based hydrodynamics on hard surfaces.

*B. Indications of surface change over the course of experiments*

We find that the patterns of $f_x$ can undergo delicate changes over a long time of repetitive experiment, even if the macroscopic control parameters remain unchanged. We speculate that the onset of stick-slip fluctuations serves as a sensitive indicator for subtle changes on the surfaces. The first indication is shown as **Fig. 6(a),** in which we present a chronological series of $f_x$ at a fixed speed $U$ with η and $D_p$ kept unchanged, over 40 repetitions of our standard protocol (Fig.2a). The stack of data shows that a gradual increase of the spatial interval that exhibits stick-slip patterns, going from a total of ~7.5mm in length initially, eventually to that of the entire sample ($L_0$).

Furthermore, we conduct similar prolonged experiments with Fig. 6(a) but with a sequence of different sliding speeds, shown as **Fig. 6(b).** The result shows that the "onset speed" at which the stick-slip pattern occurs in the sequence might also change. We find that, over the first 11 sequences of experiments, the onset speed has already shifted by almost a factor of 4. It should be noted that we have carefully prevented the change of the fluid viscosity over the long experiments: This is achieved by monitoring and compensating the concentration of fluid mixture to be within (60$\pm$1)%, so that the change of viscosity is well below 10%. Therefore, the shift of the "onset speed" cannot be attributed to the change of viscosity alone, and must be the consequence of substantial changes on the sample surface over the prolonged experiments. We speculate that the change of sample surface could be the consequence of (1) wearing over a large sliding distance, (2) the long-time absorption of molecules from the fluid, or the two factors (1) and (2) combined.

*C. Could we determine the "onset speed" of stick slips accurately?*

Whether the change between stick-slip behavior and smooth sliding is a discontinuous transition or a continuous one has been debated in literatures, and the answer often depends on the choice of tribo-pair, the normal load, and the chemical properties of the interstitial fluid [3,7,18,31,32]. One might wish to conduct a series of precisely controlled experiments with the stepping of sliding speed as small as possible, in order to pinpoint an "onset speed" --- if such transition is indeed discontinuous. However, with the analyses of Fig.6(a) and Fig.6(b), we speculate that our system might serve as an example that pinpointing the "onset speed" is practically impossible: Even if the transition does have a well-defined "onset speed", this value could also have changed over the course of our repetitive experiments. That is, the state of the system is being altered by the action of *detection* itself --- a remote analogy to the Principle of Uncertainty. This aspect also echoes our current view that, from another aspect, the change of the apparent "onset speed" might serve as a sensitive indicator for minute changes on the tribo-surfaces of concerns.

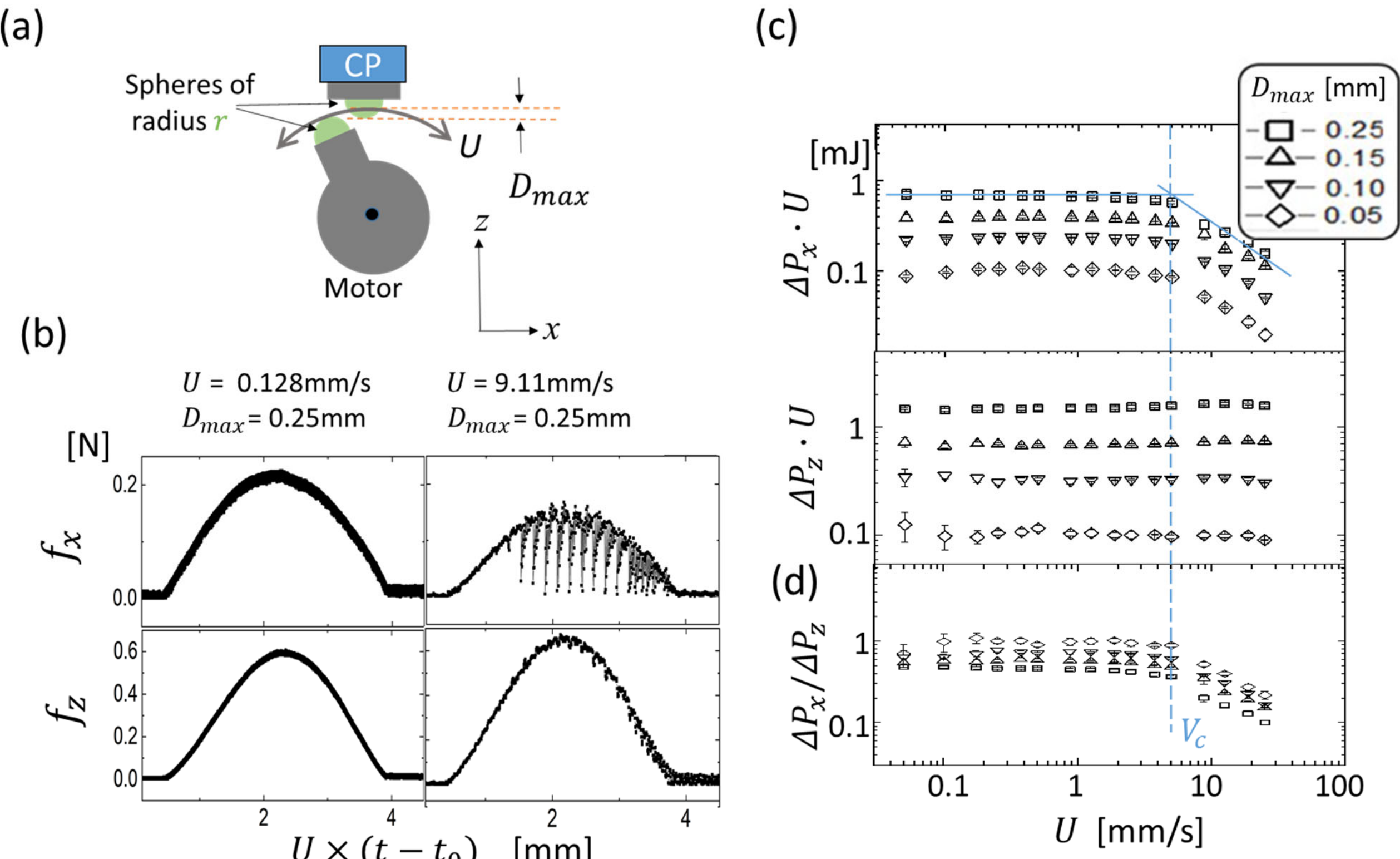

**Fig.7:** Sphere passing experiments: **(a)** The schematic setup. The dashed-lines show the the maximal overlap $D_{max}$, and the gray arrow represents the trajectory of the sample movement. **(b)** Typical time-series of tangential force $f_x$ and normal force $f_z$ for the sphere-passing events with two different sliding speeds $U$, plotted over the sliding distance $U \times (t - t_0)$ with $t_0$ being a reference time before the contact. **(c-d)** Event-integrated response $\Delta P_x \cdot U$ , $\Delta P_x \cdot U$ (defined in main text), and their ratio, as functions of the sliding speed $U$ for different values of $D_{max}$. Each symbol represents an average over five runs. The error bars represent the standard deviation over different runs. A vertical line marks the characteristic speed $V_c = 5$ mm/s as the upper boundary of the plateau, that appears to be insensitive to the value of $D_{max}$.

# V. Additional experiment beyond the fixed-depth tribology

In practical applications, solid surfaces rarely slide against each other "at a fixed pressing depth". In granular flows, for example, particles exchange momentums through events of "impact" that are neither at a fixed depth nor at a fixed force. Motivated by our prior experimental discovery of a rate-dependent intermittency in a shear flow of tightly packed PDMS spheres [27], we present in this section our additional work in investigating the force response of two spherical objects passing against each other at controlled speeds. **Fig.7(a)** shows our additional setup that has been modified for such purpose. The upper sphere is attached to the same CP as what is described in Fig.1(b). The lower sphere is driven by a motor that rotates at a constant speed back and forth, with its shaft fixed in space and pointing in the direction of *y*. We define the maximum pressing depth as $D_{max}$, in place of the $D_p$ in the main experiments, as one control parameter, while the rate of the motor controls the speed *U* at the tip. Similar to the aforementioned main experiments, both the lower sample and the contact point are fully immersed in glycerol-water mixture, but is omitted from the graph.

**Figure 7(b)** shows the time-series of the tangential force $f_x$ and the normal force $f_z$ in our two "sphere-passing events" with the same $D_{max}$=0.25mm and η ~ 10.5 mPa-s for the inteststitial fluid, but at two different values of *U*. In the case with *U*=0.128 mm/s, both $f_x$ and $f_z$ go smoothly during the sphere passing. In contrast, the case *U*=9.11 mm/s shows substantial stick slips in the signal of $f_x$. In addition, the magnitude of $f_z$ for the two cases does not show a significant change. We calculate the total momentum transfer in both direction: $\Delta P_x = \int f_x \, dt$ and $\Delta P_z = \int f_z \, dt$, respectively. Shown as **Fig.7(c-d),** both $\Delta P_x \cdot U$ and $\Delta P_z \cdot U$ show a plateau at low sliding speeds. The results, overall, are quite similar to those in our main experiments with semi-cylindrical samples, except that the curves here exhibit a relatively better defined plateau with a discontinuity in their slope, in comparison to those in Fig.3 for the fixed-depth experiments. We speculate that the reason behind this subtle difference is the gradual change of surface that we have discussed in the previous section: In the sphere-passing experiment, the total sliding distance of is short (~3 mm each run). In contrast, the total sliding distance accumulated in our fixed-depth experiments (~30mm each run) is much longer. It is understandable that a discontinuity in the slope, if any, could have been smeared out as the data is averaged over long runs of fixed-depth experiment but not for the sphere-passing events here: The sphere-passing experiment has a relatively short accumulation of sliding distance, so that the effect of wearing is not substantial.

# VI. Conclusion

We have set up an experimental system for time-resolving the vectorial force between two semi-cylindrical PDMS samples at steady sliding. Our investigations focus on the context that the contact of the two soft surfaces is fully immersed in glycerol-water mixture as a lubricant. The time-averaged tangential force reveals a non-monotonic relationship to the sliding speed, in consistence with prior literatures. For a substantial range of pressing depth ($D_P$) , we have verified that the product of the sliding speed ($U$) and the viscosity of the interstitial fluid ($\eta$, controlled by adjusting the concentration of the glycerol) indeed serves as a good parameter for characterizing the *non-monotonic* behavior: The tangential force exhibits a plateau at small values of $\eta U$, a decline over the sliding speed at intermediate values also known as the regime of *mixed lubrication* in the language of tribology, and an uprise as the speed increases at large values of $\eta U$ beyond the transition to the *hydrodynamic regime.* In contrast, we find that the normal force is insensitive to the change of sliding speed over a few decades across all three regimes: It depends only on the value of $D_P$ and appears to follow the classic Hertzian law. In addition, the results of our experiment reveal a much lower threshold for a transition to the hydrodynamic regime than the prediction by a recent theory in the context of rigid sliders [9], demanding further studies that taking into account the fluid-structure interaction between deformable surfaces.

To extract the essential tribology for modelling granular shear flows in which a large number of particle contacts are involved, we have also conducted counterpart experiments to measure the force response as two PDMS spheres slide against each other. The results reveal similar behaviors to those accomplished with two semi-cylindrical samples, and supports our interpretation of the rate-dependent intermittency already described in Ref.[27]. The identification of a well-defined characteristic speed, that is generally insensitive to the maximal pressing depth $D_{max}$, also lays the foundation of our minimal numerical model that provides further predictions [33] beyond what has been discovered in our previous work

The long travel distance and the fix-depth mechanism in our main experiments with semi-cylindrical samples allows us to analyze the time-resolved force fluctuation in response to the constant-speed sliding. Through comparing results obtained with different combinations of $U$ and $\eta$ , we find that the patterns of stick-slip fluctuations, that is prominent only in the mixed-lubrication regime, are roughly controlled by the product $\eta U$. We find signs indicating that prolong, repetitive experiments could have changed the surface of our samples substantially: The change is reflected by the gradual change of the stick-slip behaviors at the vicinity of its onset. Although this phenomenon might have prevented us from identifying an "onset speed" accurately, we believe that monitoring the occurrence of stick slip may find its potential application in detecting delicate changes of a soft surface that keeps evolving.

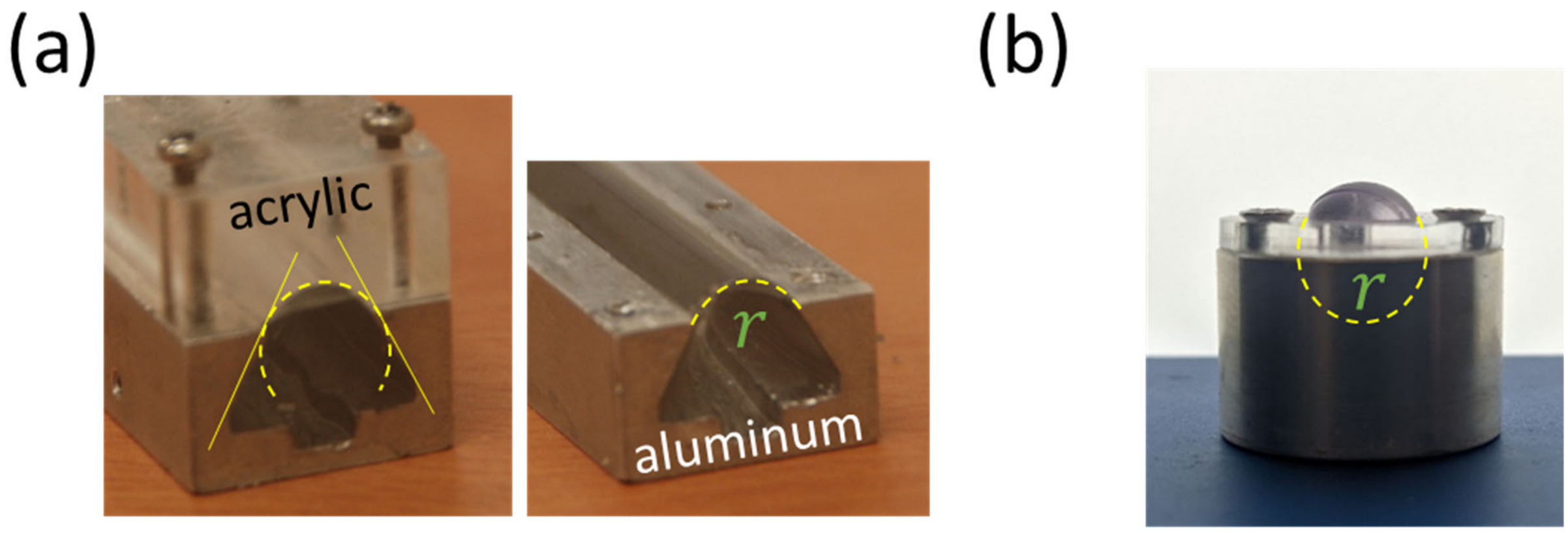

**Fig. A1:** Photographs of our PDMS sample, all taken by a standard single-lens camera in room light. **(a)** The aluminum base of the sample, with and without the acrylic mold, respectively. The curvature of the sample surface along the transverse direction is $r = 4.5\text{mm}$ . **(b)** Assembly of our spherical sample and its aluminum base. The base has a semi-spherical indent (shown by the dashed line) that fits the sample. An acrylic ring with a tapered inner edge and flushed screws are used to keep the sphere from moving.

# Appendix

*Molding of the semi-cylindrical sample* --- Our semi-cylindrical PDMS samples are prepared by molding. The mold for shaping the sample is an assembly consisting of an aluminum base and an acrylic enclosure, shown as Fig.A1(a). Both are precision machined to the accuracy of 0.01 mm. Given such precision of both, the 10 cm-long upper surface of the aluminum base is parallel to the "ridge" of the molded PDMS surface within the same error, and therefore serve as a plane of reference in checking our alignment of the setup --- see below for further description.

*Alignment of the driving mechanism* ---To ensure the pressing depth $D_p$ is strictly a constant throughout the centimeter-long travel of the lower sample every time, great cares are taken in alighing the ridge of the PDMS surface with the direction of motion for the motorized translational statge. This is achieved by pointing a laser displacement sensor (Optex CD33-50NV, at a precision of 0.002 mm) to the reference plane defined above for the lower sample, while the translational stage is set into motion. We finetune junctions between the lower sample, the container, and the motorized stage and have verified that, the vertical displacement of the laser spot in relation to the rest of the assembly (where the upper sample is firmly attached) is much less than 0.01 mm. These procedures not only defines the "direction of x" accurately, but also ensures that the value of $D_p$, once set, can remain constant well within the error described above throughout the travel of the lower sample for each run

*Choice of sensors*---We have chosen sensors that are much stiffer than the PDMS. Sensor S0 has the lowest stiffness, but still ensures that displacement of the CP (where the upper sample is attached) in response to the forcing between two sample is always much smaller than the set value of $D_p$ by a factor of 1/20. Under such condition, the inertia of the measurement system (except the CP) plays no roles in the time-dependent force signals recorded. We have also assessed the potential contribution from the acceleration of the upper sample and CP can be safely ignored.

*Gelation of the elastomer*--- Both the spherical and semi-cylindrical samples undergo the same protocol for gelation. The mold is fully immersed in a fluidic mixture made of two agents, at a ratio of 10 to 1 as recommended by the manufacturer [34] to reach the specified elasticity. We degas the mixture by reducing the ambient pressure (to ~0.1atm) for a few hours, before baking it at 70 degrees Celsius overnight to ensure a complete gelation. The Young's modulus of both samples is determined by independent experiments described previously in the Appendix-C of Ref.[26] to be ~1.5 MPa, in consistence with prior studies [35,36].

*Spherical sample* --- We use the 9mm-diameter PDMS spheres that we have produced previously for prior experiments. The spherical samples are also prepared by molding with the same gelation procedure of making the semi-cylindrical sample. See Appendix-C in Ref. [27] for the detail of production. An image of the spherical sample on the adaptor is shown as **Fig.A1(b).**